\documentclass[prd,showpacs,preprintnumbers,amsmath,amssymb]{revtex4}

\usepackage{graphicx}
\usepackage{dcolumn}
\usepackage{bm}

\begin{document}

\title{Schwinger Pair Production in Electric and Magnetic Fields}

\author{Sang Pyo Kim}\email{sangkim@kunsan.ac.kr;spkim@phys.ualberta.ca}

\affiliation{Department of Physics, Kunsan National University,
Kunsan 573-701, Korea} \affiliation{Asia Pacific Center for
Theoretical Physics, Pohang 790-784, Korea}

\author{Don N. Page}\email{don@phys.ualberta.ca}

\affiliation{Theoretical Physics Institute, Department of Physics,
University of Alberta, Edmonton, Alberta, Canada T6G 2J1}

\date{\today}

\begin{abstract}
Charged particles in static electric and magnetic fields have Landau levels and
tunneling states from the vacuum. Using the instanton method of Phys. Rev. D
{\bf 65}, 105002 (2002), we obtain the formulae for the pair-production rate in
spinor and scalar QED, which sum over all Landau levels and recover exactly the
well-known results. The pair-production rates are calculated for an electric
field of finite extent, and for the Sauter potential, both with a constant
magnetic field also present, and are shown to have finite-size effects.
\end{abstract}
\pacs{PACS number(s): 12.20.-m, 13.40.-f}

\maketitle

\section{Introduction}

Vacuum polarization and pair production are two physically important
phenomena of quantum electrodynamics (QED) in strong electromagnetic
fields \cite{heisenberg,schwinger}. The one-loop effective action in
constant electric and magnetic fields has nonlinear contributions to
the classical action and, thereby, leads to the nonlinear Maxwell
equations. Another interesting phenomenon is the pair production due
to vacuum instability in the presence of electric fields near and
above the critical electric field strength $E_c = m^2 c^3/e \hbar
~(1.3 \times 10^{16}\, \mathrm{V/cm})$ \cite{heisenberg,schwinger}
(for references, see, e.g., \cite{kim-page,dunne}).

Strong QED has many physical applications \cite{greiner}. In
particular, electromagnetic fields of some neutron stars and black
holes \cite{putten} and high-intensity laser fields \cite{burke}
above the critical strength $E_c /c = B_c = m^2 c^2/e \hbar ~ (4.4
\times 10^{13} G)$ have revived recent interest and applications of
strong QED. The one-loop QED effective action is known exactly in
the background of constant electric and magnetic fields
\cite{schwinger}. In the case in which ${\bf E}^2 - {\bf B}^2$ and
${\bf E} \cdot {\bf B}$ are not both zero, one can go to a frame in
which ${\bf E}$ and ${\bf B}$ are parallel with magnitude $E$ and
$B$, and obtain the imaginary part of the one-loop effective action
per four-volume for spinor QED given by
\cite{heisenberg,schwinger,bunkin-tugov,daugherty}
\begin{equation}
2 {\rm Im} {\cal L}^{(1)}_{\rm spinor} = w_{\rm spinor}^{(1)} =
\frac{(qE)(qB)}{ (2 \pi)^2} \sum_{n = 1}^{\infty} \frac{1}{n} {\rm
coth} \Bigl(\frac{n \pi B}{E} \Bigr) \exp\Bigl(- \frac{n \pi
m^2}{qE} \Bigr), \label{form2}
\end{equation}
and that for scalar QED by \cite{heisenberg,popov,itzykson,cho}
\begin{equation}
2 {\rm Im} {\cal L}^{(1)}_{\rm scalar} = w_{\rm scalar}^{(1)} =
\frac{(qE)(qB)}{2 (2 \pi)^2}
\sum_{n = 1}^{\infty} \frac{(-1)^{n+1}}{n}
{\rm csch} \Bigl( \frac{n \pi B}{E} \Bigr)
\exp \Bigl(- \frac{n \pi m^2}{qE} \Bigr). \label{form1}
\end{equation}
The pair-production rates per volume per time
themselves are the first terms in each of these
series \cite{nikishov},
\begin{equation}
{\cal N}^{(1)}_{\rm spinor} = \frac{(qE)(qB)}{(2 \pi)^2}
{\rm coth} \Bigl(\frac{\pi B}{E} \Bigr)
\exp\Bigl(- \frac{\pi m^2}{qE} \Bigr), \label{fer pair}
\end{equation}
and
\begin{equation}
{\cal N}^{(1)}_{\rm scalar} = \frac{(qE)(qB)}{2 (2 \pi)^2}
{\rm csch} \Bigl( \frac{\pi B}{E} \Bigr)
\exp \Bigl(- \frac{ \pi m^2}{qE} \Bigr). \label{bos pair}
\end{equation}
Kruglov derived a general formula for the more general case of
particles of arbitrary spin $s$ with ADM (anomalous magnetic
moment) and EDM (electric dipole moment) \cite{kruglov}.

In this paper we apply the recently-introduced instanton method
\cite{kim-page} to find the pair-production rate
in constant electric and magnetic
fields in spinor and scalar QED, agreeing with the results above.
The idea of the instanton method,
the elaboration of the role of tunneling in pair production
\cite{brezin}, is that for static electric fields, fermions and
bosons have tunneling states from the vacuum in the Coulomb
(space-dependent) gauge, and their pair-production rates are
determined by these instanton actions for tunneling. In the
presence of static magnetic fields, charged fermions and bosons
have discrete spectrum of Landau levels. Taking into account both
Landau levels and instanton actions for tunneling, we obtain
equivalent formulae for pair-production rates in spinor and scalar
QED. These formulae are given as sums over all Landau levels, and
we recover exactly the well-known results. Further, applying
the method, we estimate the pair-production rates by an electric
field with a finite extent and a constant magnetic field, which
have additional factors determined by the potential difference
across the boundary beside the pair-production rate for a constant
electromagnetic field.

The organization of this paper is as follows. In Sec. II we
generalize the instanton method to the case of constant electric
and magnetic fields and obtain the pair-production rate as the sum
of Landau levels. We also compare our instanton method with the
instanton method in Euclidean time. In Sec. III
 we apply the instanton method to
inhomogeneous electric fields: a constant electric field confined to a
finite region and one from the Sauter-type potential.

\section{Instanton Method for Particles with Spin in Electromagnetic Fields}

With the spin properly taken into account, the component equation
for minimally-coupled particles with spin
in a constant electric field takes the form
(in units with $\hbar = c =1$ and with metric signature $(+, -, -, -)$)
\begin{equation}
\Bigl[\eta^{\mu \nu} (\partial_{\mu} + i q A_{\mu})
(\partial_{\nu} + i q A_{\nu}) + m^2 + 2 i \sigma q E \Bigr]
\Phi_{\sigma} = 0, \label{vec eq}
\end{equation}
where  $q~ (q > 0)$, $m$ and $\sigma$ are the charge,
 mass and spin of the particles. We shall show the
physical implication of the imaginary part from the spin effect.
After mode-decomposition,
\begin{equation}
\Bigl[- \partial_z^2 - (\omega + qEz)^2 + m^2 + {\bf k}^2_{\perp}
+ 2 i \sigma qE \Bigr] \Phi_{\sigma \omega {\bf k}_{\perp}} (z) =
0. \label{comp eq}
\end{equation}
The wave function (\ref{comp eq}) has a solution in terms of the
complex parabolic cylindrical function $E(x,y)$ \cite{kim-page,e-m sum},
\begin{equation}
\Phi_{\sigma \omega {\bf k}_{\perp}} (z) = C E(\tilde{a}_{s {\bf
k}_{\perp}}, \xi),
\end{equation}
where
\begin{equation}
\xi = \sqrt{\frac{2}{qE_0}} (\omega + q E z), \quad \tilde{a}_{s
{\bf k}_{\perp}} = \frac{m^2 + {\bf k}_{\perp}^2}{2 qE} + i
\sigma.
\end{equation}
In the two asymptotic regions the wave function becomes
\begin{eqnarray}
\Phi_{\sigma \omega {\bf k}_{\perp}} (z) &=& A \varphi_{\sigma
\omega {\bf k}_{\perp}} (\xi) - B \varphi^*_{\sigma \omega {\bf
k}_{\perp}} (\xi), \quad (\xi \ll - 2 |\tilde{a}_{s {\bf
k}_{\perp}}|^{1/2}), \nonumber\\ \Phi_{\sigma \omega {\bf
k}_{\perp}} (z) &=& C \varphi^*_{\sigma \omega {\bf k}_{\perp}}
(\xi), \quad (\xi \gg  2 |\tilde{a}_{s {\bf k}_{\perp}}|^{1/2}),
\end{eqnarray}
where
\begin{equation}
\varphi_{\sigma
\omega {\bf k}_{\perp}} (\xi) = \sqrt{\frac{2}{|\xi|}} e^{- (i/4) \xi^2},
\end{equation}
and
\begin{equation}
A = i C ( 1 + e^{2 \pi \tilde{a}_{s {\bf k}_{\perp}}})^{1/2}, \quad
B = - i C e^{\pi \tilde{a}_{s {\bf k}_{\perp}}}.
\end{equation}

For bosons $(\sigma = 0, 1, 2, \cdots)$, we have the flux
conservation
\begin{equation}
|A|^2 = |B|^2 + |C|^2.
\end{equation}
The reflection probability is given by
\begin{equation}
\Bigl|\frac{B}{A} \Bigr|^2 = \frac{e^{ \pi ( \tilde{a}^*_{s {\bf
k}_{\perp}} + \tilde{a}_{s {\bf k}_{\perp}})}}{ [( 1+ e^{ 2 \pi
\tilde{a}^*_{s {\bf k}_{\perp}}})(1 + e^{ 2 \pi  \tilde{a}_{s {\bf
k}_{\perp}}})]^{1/2}} = \frac{1}{1 + e^{- 2 S_{{\bf k}_{\perp}}}},
\end{equation}
where $S$ is the instanton action
\begin{equation}
S_{{\bf k}_{\perp}} = \pi \frac{m^2 + {\bf k}_{\perp}^2}{2 qE}.
\end{equation}
This is with the boundary condition that the current flux on the
right hand side is outward. However, then the group velocity there
is inward (to the left). If we instead impose the boundary
condition from causality that signals are outgoing on the right,
then on the left hand side the outgoing flux becomes $|A/B|^2 = 1
+ e^{- 2S}$ times the incoming flux, an amplification by the
Klein-paradox. Then by the results of Nikishov \cite{nikishov},
the pair-production rate is just this amplification factor minus
1, or $e^{-2S}$. It is interesting that the instanton
approximation with positive action $S$ (exact for a uniform field)
never gives more than 1 expected pair per mode, though there is no
such restriction for a general electric field.

On the other hand, for fermions $(\sigma = 1/2, 3/2, \cdots)$
there is the Klein-paradox, and the flux conservation now becomes
\begin{equation}
|A|^2 +|C|^2 = |B|^2.
\end{equation}
This is also with outgoing flux but incoming signals on the right. Replacing
this by outgoing signals on the right makes the outgoing flux on the left
become $|A/B|^2 = 1 - e^{-2S}$ times the incident flux on the left. (There is
no amplification factor larger than one in the fermion case with causality
imposed, because of the Pauli exclusion principle, as Feynman once explained
to one of us (DNP) while drawing diagrams and saying, ``I'm supposed to be
good at these diagrams.'') Then by Nikishov's results \cite{nikishov},
the pair-production rate is just one minus this reflection coefficient,
again $e^{-2S}$.
This result confirms the use of the instanton method in this paper
and \cite{kim-page}.

In the instanton method \cite{kim-page}, we use  the
Klein-Gordon equation
\begin{equation}
\Bigl[ \eta^{\mu \nu} (\partial_{\mu} + i
q A_{\mu}) (\partial_{\nu} + i q
A_{\nu})  + m^2 \Bigr] \Phi (t, {\bf x}) = 0. \label{kg
eq}
\end{equation}
The significant difference between Eqs. (\ref{vec eq}) and (\ref{kg eq})
is the imaginary constant term. We illustrate the instanton method by
first considering the case of a pure electric field along the $z$
direction. In the Coulomb gauge, the 4-potential is given by
\begin{equation}
A_{\mu} = (-Ez, 0, 0, 0).
\end{equation}
The component field of the Klein-Gordon equation,
\begin{equation}
[(\partial_t - i q E z)^2 - \partial_x^2 - \partial_y^2
- \partial_z^2 + m^2 ] \Phi = 0,
\end{equation}
has a solution of the form
\begin{equation}
\Phi = e^{ i ({\bf k}_{\perp} \cdot {\bf x}_{\perp} - \omega t)}
\Phi_{\omega {\bf k}_{\perp}} (z),
\end{equation}
where ${\bf k}_{\perp}$ and
${\bf x}_{\perp}$ denote the momentum and the vector perpendicular to
the electric field. Then the above equation becomes
\begin{equation}
[ - \partial_z^2
- (\omega +  q E z)^2 + m^2 + {\bf k}_{\perp}^2 ]
\Phi_{\omega {\bf k}_{\perp}} (z) = 0. \label{red eq}
\end{equation}
Note that Eq. (\ref{red eq}) describes the wave equation for an inverted
harmonic potential with a negative energy. Therefore,
one has tunneling states under the potential barrier,
whose instanton actions are given by \cite{kim-page}
\begin{equation}
2 S_{{\bf k}_{\perp}} = \oint Q^{1/2} (z) = \pi
\frac{m^2 + {\bf k}_{\perp}^2}{qE},
\end{equation}
where $Q(z) = m^2 + {\bf k}_{\perp}^2 - (\omega +  q E z)^2$. The tunneling
state and instanton action can also be understood in classical theory.
The mass of the charged particle is invariant:
\begin{equation}
m^2 = \eta^{\mu \nu} (p_{\mu} - q A_{\mu})
(p_{\nu} - q A_{\nu}) = (p_0 + qEz)^2 - p_z^2 - {\bf p}_{\perp}^2.
\label{mass}
\end{equation}
The time component and the transverse component of 4-momentum are
constants of motion, say, $p_0 = \omega$ and ${\bf p}_{\perp}
= {\bf k}_{\perp}$. In the Euclidean spacetime $t = i \tau $,
the $z$ component
becomes $p_z = i p_z^{E}$, where $p_z^{E}$ is the Euclidean momentum.
Hence Eq. (\ref{mass}) becomes a harmonic
oscillator in the reduced phase space $(z, p_z)$,
\begin{equation}
\frac{1}{2} p_z^2 + \frac{(qE)^2}{2}(z + \frac{\omega}{qE} )^2
= \frac{1}{2}(m^2 + {\bf k}_{\perp}^2).
\end{equation}
The action of the oscillator, given by the energy divided by
the frequency $(qE)/ (2 \pi)$, is the instanton action
\begin{equation}
{\cal S}_{{\bf k}_{\perp}} = 2 S_{{\bf k}_{\perp}}
 = \pi \frac{m^2 + {\bf k}_{\perp}^2 }{qE}. \label{inst action}
\end{equation}
The instanton action (\ref{inst action}) can also be obtained in
the time-dependent gauge $A_{\mu} = (0,0,0, -Et)$. Using the
action ${\cal S} = \int L(t, {\bf x})$, Popov \cite{popov}
obtained the same instanton action in the Coulomb gauge, and
Affleck et al. \cite{affleck} also obtained the same result in the
mixed-gauge $A_{\mu} = - F_{\mu \nu} x_{\nu}/2$. Therefore, the
instanton action does not depend on the choice of gauge, which is
just a matter of technical simplicity.

In a pure magnetic field ${\bf B} = B {\bf e}_z$ along the z-direction,
${\bf A} = (0, Bx, 0)$, and the component field equation of the Klein-Gordon
or Dirac equation has a solution of the form
\begin{equation}
\Phi = e^{ i (k_y y + k_z z - \omega t)}
\Phi_{\sigma_{\pm} \omega k_y k_z}(x),
\end{equation}
which leads to the mode equation
\begin{equation}
\Bigl[- \partial_x^2 + (q Bx + k_y)^2
 + m^2 + k_z^2 - 2 qB \sigma_{\pm} - \omega^2 \Bigr]
\Phi_{\sigma_{\pm} \omega k_y k_z}(x) = 0. \label{ks osc}
\end{equation}
Here  $\sigma_{\pm}$ is the spin projection: $\sigma = 0$ for the
scalar and $\sigma_{\pm} = \pm 1/2$ for the Dirac spinor. The mode
equation (\ref{ks osc}) has bound states given by harmonic wave
functions with the energy spectrum \cite{greiner2}
\begin{equation}
\omega^2 = m^2 + k_z^2 + qB (2j+ 1 - 2 \sigma_{\pm}),
\quad (j = 0,1, \cdots). \label{sc lan}
\end{equation}
The discrete spectrum due to the magnetic field is the Landau levels for
charged particles. Note that all the Landau levels are non-degenerate for
the scalar particles, whereas all the states of the Dirac spinor
are doubly degenerate, $j, \sigma_+ = 1/2$ and $j-1, \sigma_- = - 1/2$,
except for the unique lowest Landau level, $j = 0, \sigma_+ = 1/2$.

Now we apply the formulae \cite{kim-page} for pair production based
on the instanton calculation to the static uniform electric and magnetic
fields. For the electric and magnetic fields parallel
to each other along the z-direction, the 4-potential is given by
\begin{equation}
A_{\mu} = (-Ez, 0, Bx, 0).
\end{equation}
The component field equation,
\begin{equation}
[(\partial_t - i q E z)^2 - \partial_x^2 - (\partial_y + i qBx)^2
- \partial_z^2 + m^2 - 2qB \sigma_{\pm} ] \Phi_{\sigma_{\pm}} = 0,
\end{equation}
has the solution of the form
\begin{equation}
\Phi_{\sigma_{\pm}} = e^{ i (k_y y - \omega t)} \Phi_{\sigma_{\pm} \omega k_y} (x,z).
\end{equation}
Then the above equation becomes
\begin{equation}
[\{- \partial_x^2 + (qBx+ k_y)^2\}  + \{- \partial_z^2
- (\omega +  q E z)^2 \} + m^2 - 2 qB \sigma_{\pm} ]
\Phi_{\sigma_{\pm} \omega k_y} (x, z) = 0.
\end{equation}
The first parenthesis has the harmonic wave functions as
eigenfunctions, so the remaining equation becomes
\begin{equation}
[- \partial_z^2 -
(\omega +  q E z)^2 + m^2 + q B (2j+1 - 2 \sigma_{\pm}) ]
\Phi_{\sigma_{\pm} \omega k_y j} (z) = 0,
\quad (j = 0, 1, \cdots, ).
\end{equation}
Therefore, the charged particles, exactly described by
the inverted harmonic potential, have, as
the instanton action for tunneling states,
\begin{equation}
2 S_{\sigma_{\pm} j} = \pi \frac{m^2 + qB
(2 j+1 - 2 \sigma_{\pm})}{qE}. \label{inst1}
\end{equation}

The main result of  Ref. \cite{kim-page}, when
converted from the imaginary part of the effective action to
the pair-production, is that the expected number of pairs produced
per mode in a static electric field is given by
\begin{equation}
N_{\pm} =  \pm (e^{\pm w} -1)  = e^{-2S}, \label{33}
\end{equation}
where the upper (lower) sign is for bosons (fermions)
and $S$ is the instanton action for the corresponding mode, here
given in Eq. (\ref{inst1}). We then obtain the pair-production
rate \cite{footnote} for scalar particles as
\begin{equation}
{\cal N}_{\rm scalar}^{(1)} (E, B, m)
=  \frac{(qE)(qB)}{(2 \pi)^2}
 \sum_{j = 0}^{\infty} \exp \Bigl(-  \pi \frac{m^2
+ qB (2 j+1)}{qE} \Bigr), \label{b-p}
\end{equation}
and for fermions as
\begin{equation}
{\cal N}_{\rm spinor}^{(1)} (E, B, m) = \frac{(qE)(qB)}{(2 \pi)^2}
\sum_{j = 0}^{\infty} \sum_{\sigma_{\pm} = \pm 1/2}
\exp \Bigl(- \pi \frac{m^2
+ qB (2 j+1 - 2 \sigma_{\pm})}{qE} \Bigr). \label{f-p}
\end{equation}
Here $(qE)/(2 \pi)$ is the (number) density of states (per momentum)
from the $\omega$-integration, and $(qB)/(2 \pi)$ is that available
for each Landau level. When we sum the geometric series of Eqs.
(\ref{b-p}) and (\ref{f-p}), we readily get the standard results of
Eqs. (\ref{fer pair}) and (\ref{bos pair}) respectively.

It should be noted that the pair-production rate
does not depend on the renormalization scheme, since
all divergence and renormalizability is contained in ${\rm Re}
{\cal L}$ \cite{schwinger,kruglov}.

\section{Inhomogeneous Field}

We now extend the analysis to inhomogeneous electric fields together
with a constant magnetic field. As inhomogeneous electric fields
we consider a localized constant electric field in
the region $\vert z \vert \leq L$
and an electric field obtained from the Sauter potential $
A_0 (z) = - E L \tanh (z / L)$ for $qEL \gg m$ and $mL \gg 1$.
In both cases, the electric
field extends effectively a distance $2L$ in the $z$-direction.
For an electric field localized in the $z$-direction,
pairs are produced only when $\omega -
qA_0 (+ \infty) \geq m$ and $\omega - qA_0 (- \infty) \leq - m$. Also
the instanton actions exist when the Landau levels are limited to
\begin{equation}
qB (2 j_{\rm max} + 1) = {\rm min} \{ (\omega - qA_0(+ \infty))^2 - m^2,
(\omega - qA_0 (- \infty))^2 - m^2 \}.
\end{equation}
The pair-production rate per area and per time (using an overbar to
distinguish from rates per volume per time)
now takes the form, for scalar QED,
\begin{eqnarray}
\overline{\cal N}_{\rm scalar}^{(1)} = \frac{(qB)}{(2\pi)^2}
 \int_{qA_0(+ \infty) + m}^{qA_0 (- \infty) -m} d \omega
\sum_{j = 0}^{j_{\rm max}} e^{ - 2S_j},  \label{inh bp}
\end{eqnarray}
and for spinor QED
\begin{eqnarray}
\overline{\cal N}_{\rm spinor}^{(1)} = \frac{(qB)}{(2 \pi)^2}
 \int_{qA_0 (+ \infty) + m}^{qA_0 (- \infty) -m} d \omega
\sum_{j = 0}^{j_{\rm max}} \sum_{\sigma_{\pm} = \pm 1/2}
e^{ - 2S_{\sigma_{\pm} j} }  , \label{inh fp}
\end{eqnarray}
where $(qB)/(2\pi)$ is the number of state for Landau levels and
another factor $1/(2\pi)$ is from the $\omega$ integration.
The $\omega$ integration yields the potential energy difference
\begin{equation}
V = qA_0(- \infty) - q A_0
(+ \infty) = 2 qE L,
\end{equation}
where $L$ would extend to infinity for the homogeneous electric field.
Here the instanton actions are determined by
\begin{equation}
S_{\sigma_{\pm} j} = \sum_{k = 0}^{\infty} S^{(2k)}_{\sigma_{\pm} j},
\end{equation}
where the dominant (0-loop or classical) contribution comes from
\begin{equation}
S^{(0)}_{\sigma_{\pm} j} = \int_{z_-}^{z_+} dz \sqrt{m^2 + qB (2j + 1 -
2 \sigma_{\pm}) -
(\omega - q A_0 (z))^2},
\end{equation}
where $z_{\pm}$ are turning points of the integral and $\sigma =
0$ for scalar and $\sigma_{\pm} = \pm 1/2$ for spinor QED.

In the first case the electric field is confined to a finite region
of length $2L$. The potential is given by
\begin{equation}
A_{\mu} = (- Ez, 0, Bx, 0), \quad \vert z \vert \leq L.
\end{equation}
The instanton actions are given by
\begin{equation}
2 S_{j} = \pi \frac{m^2 + qB (2 j+1)}{qE}
\end{equation}
for $j$ smaller or equal to the highest Landau level
\begin{equation}
j_{\rm max} = \frac{1}{2qB} \Biggl[\Bigl( \frac{V}{2} - |\omega| \Bigr)^2
- m^2 \Biggr] - \frac{1}{2}. \label{high lan}
\end{equation}
Finally, after the $\omega$ integration, we obtain the pair-production
rate per area and per time for scalar QED
\begin{eqnarray}
\overline{\cal N}_{\rm scalar}^{(1)} &=& \frac{(qE) (qB)}{2(2\pi)^2}
  {\rm csch} \Bigl(\frac{\pi B}{E} \Bigr)
\exp \Bigl( - \frac{\pi m^2}{qE}\Bigr) \nonumber\\
&& \times (2L) \Biggl[ 1 - \frac{2m}{V}
+ \frac{2m}{V} \exp \Bigl(- \frac{\pi B}{E} \Bigr) \int_0^{\sqrt{
(V/2m)^2 -1}} dx \frac{x e^{- \frac{\pi m^2}{qE} x^2}}
{\sqrt{1 +x^2}} \Biggr], \label{step1}
\end{eqnarray}
and that for spinor QED
\begin{eqnarray}
\overline{\cal N}_{\rm spinor}^{(1)} &=& \frac{(qE) (qB)}{(2\pi)^2}
 {\rm coth} \Bigl(\frac{\pi B}{E} \Bigr)
\exp \Bigl( - \frac{\pi m^2}{qE}\Bigr) \nonumber\\
&& \times (2L) \Biggl[ 1 - \frac{2m}{V}
+ \frac{2m}{V} \exp \Bigl(- \frac{\pi B}{E} \Bigr) \int_0^{
\sqrt{(V/2m)^2 -1}} dx \frac{x e^{- \frac{\pi m^2}{qE} x^2}}
{\sqrt{1 +x^2}} \Biggr]. \label{step2}
\end{eqnarray}
We note that the first factor is the pair-production rate per volume for
the constant electric field in Sec. II, and
the localization of the electric field gives the factor on the second line,
which is roughly $2L$ for $V \gg 2m$.
In the large $L$ limit, after dividing by $2L$ to convert from a rate
per area to a rate per volume, we recover Eqs. (\ref{fer pair}) and
(\ref{bos pair}), since $m/V \rightarrow 0$.

Next, we turn to the slowly varying electric field $E(z) = E
{\rm sech}^2 (z/L)$
given by the Sauter potential
\begin{equation}
A_0 (z) = - EL \tanh \Bigl(\frac{z}{L} \Bigr),
\end{equation}
for $qEL \gg m$ and $mL \gg 1$. The highest Landau level
is the same as Eq. (\ref{high lan}).
However, the instanton actions are given by
\begin{equation}
2 S_{\sigma_{\pm} j} = \frac{\pi m^2 \alpha}{qE}  + \frac{\pi B \beta}{E}
(2 j + 1 - 2 \sigma_{\pm}) + {\cal O}
\Bigl(\frac{j^2}{V^2}, \frac{1}{V^4} \Bigr),
\end{equation}
where
\begin{eqnarray}
\alpha &=& 1 + 4 \frac{\omega^2}{V^2}+ \frac{m^2}{V^2}, \nonumber\\
\beta &=& 1 + 4 \frac{\omega^2}{V^2} + 2 \frac{m^2}{V^2}.
\end{eqnarray}
The instanton actions are also obtained from Eq. (47) of Ref.
\cite{kim-page} by replacing ${\bf k}^2_{\perp}$
by $qB (2 j + 1 - 2 \sigma_{\pm})$. Then the pair-production rate
per area and per time for scalar QED takes the form
\begin{eqnarray}
\overline{\cal N}_{\rm scalar}^{(1)} &=& \frac{(qB)}{2 (2\pi)^2}
\int_{- (\frac{V}{2} - m)}^{(\frac{V}{2} -m)}
d \omega  ~{\rm csch} ~\Bigl(\frac{\pi B \beta}{E} \Bigr)
\exp \Bigl( - \frac{\pi m^2 \alpha}{qE}\Bigr)\nonumber\\
&& \times
\Biggl[ 1- \exp \Bigl( - \frac{\pi B \beta}{E} \Bigr)
\exp \Bigl\{ - \frac{\pi \beta}{qE} \Bigl( (\frac{V}{2} - |\omega|)^2
- m^2 \Bigr) \Bigr\} \Biggr]. \label{sauter}
\end{eqnarray}
The spinor case is obtained by replacing ${\rm csch}(\pi B \beta/E)$
by $2 {\rm coth}(\pi B \beta/ E)$, the 2 being for two spins.
In the $L = \infty$ limit, with $\int d \omega = V = 2 qEL = \infty$,
we recover the standard result (\ref{fer pair}) and (\ref{bos pair})
when we divide by $2L$ to get the rate per volume instead of the
rate per area. In fact, we can see that the final factor of the
right hand side of Eq. (\ref{sauter}), the part inside the square brackets,
is nearly one over most of the integral if $qEL^2 \gg 1$, so that that
factor can be dropped.

It is interesting to compare the pair-production rates per area given by
the instanton method using just the classical action
with the rate per area given by integrating the
uniform-field rate per volume over the extent of the inhomogeneous field:
\begin{equation}
\overline{\cal N}_{\rm u} = \int dz ~{\cal N} (E(z), B, m).
\end{equation}
Here the subscript u is for ``uniform field''.

For simplicity, let us take the case $B = 0$, so the uniform-field rate
per volume is the same for bosons and for fermions (per spin state)
and in the Sauter
electric field $E(z) = E {\rm sech}^2(z/L)$ would be
\begin{equation}
{\cal N} = \frac{(qE)^2}{(2 \pi)^3} {\rm sech}^4 \Bigl(\frac{z}{L} \Bigr)
\exp \Bigl[- \frac{\pi m^2}{qE} \cosh^2 \Bigl(\frac{z}{L} \Bigr)
\Bigr].
\end{equation}
If we let
\begin{equation}
\delta \equiv \frac{qE}{\pi m^2}
\end{equation}
(where this $E$ is the maximum value of the Sauter electric field),
and if we let $x = \sinh (z/L)$, we get
\begin{equation}
\overline{\cal N}_{\rm u} = \frac{(qE)^2 L}{(2 \pi)^3} e^{- \frac{1}{\delta}}
\int_{- \infty}^{\infty} \frac{dx}{(1+ x^2)^{5/2}} e^{- \frac{x^2}{\delta}}.
\end{equation}

Analogously, if in Eq. (\ref{sauter}) we set $x = 2 \omega/V$ and
\begin{equation}
\epsilon \equiv \frac{2m}{V} = \frac{m}{qEL} \ll 1,
\end{equation}
then assuming $\delta \epsilon^2 = 1 /(\pi q E L^2) \ll 1$ so that the final
factor in Eq. (\ref{sauter}) may be dropped, that equation gives
\begin{equation}
\overline{\cal N}_{\rm i} \approx  \frac{(qE)^2 L}{(2 \pi)^3} e^{- \frac{1}{\delta}}
\int_{- 1}^{1} \frac{dx}{1+ x^2} e^{- \frac{x^2}{\delta}}.
\end{equation}
Here the subscript i stands for ``instanton method'', here using just
the 0-loop or classical action.

If we now have a maximum electric field $E$ far below the critical value
$m^2/q$, so $\delta \ll 1$, then
\begin{equation}
\overline{\cal N}_{\rm u} \approx e^{\frac{\epsilon^2}{4 \delta}}
\overline{\cal N}_{\rm i} \approx  \frac{(qE)^2 L}{(2 \pi)^3}
\sqrt{\pi \delta} e^{- \frac{1}{\delta}} =
\frac{(qE)^{5/2} L}{(2 \pi)^3 m} e^{- \frac{1}{\delta}}. \label{in res}
\end{equation}
The two estimates for the rate agree if
\begin{equation}
\frac{\epsilon^2}{4 \delta} = \frac{\pi m^4}{4 (qE)^3 L^2}
= \frac{\pi m^4}{qE V^2} = \frac{2 \pi m^4 L}{V^3} \ll 1,
\end{equation}
but they do not agree if this quantity is not small.

On the other hand, if the maximum electric field $E$ is far above
the critical value $m^2/q$, so $\delta \gg 1$, then necessarily
$\epsilon^2 / \delta \ll 1$, and we get
\begin{equation}
\overline{\cal N}_{\rm u} \approx \frac{8}{3 \pi}
\overline{\cal N}_{\rm i} \approx  \frac{(qE)^2 L}{6 \pi^3}
\sqrt{\pi \delta}. \label{un res}
\end{equation}
Thus in this case the uniform-field approximation gives roughly
$85\%$ of the 0-loop instanton-method approximation using only the leading term
for the actions.

It is also of interest to compare these results with
the exact results for the Sauter electric field. Nikishov \cite{nikishov}
has given the expected number of pairs per mode, but one must integrate
over the modes to get the pair-production rate per area. Here we shall
restrict ourselves to spinor case, as it is actually simpler. Then
if one defines
\begin{equation}
Z \equiv 2 \pi qEL^2 = \frac{2}{\delta \epsilon^2},
\end{equation}
the pair-production rate per area for spinors (with two spin states) can be
shown after some algebra to be
\begin{equation}
\overline{\cal N}_{\rm spinor} = 2 \frac{(qEL)^3}{(2 \pi)^2}
\int \frac{\cosh Zy - \cosh Zx}{\cosh Z - \cosh Zx} (y^2 - x^2) dx dy,
\label{i}
\end{equation}
where in this integral we have the restrictions
\begin{eqnarray}
- 1 \leq - y \leq x \leq y \leq 1, \nonumber\\
(1 - x^2) (1 - y^2) \geq \Bigl(\frac{2m}{V} \Bigr)^2 = \epsilon^2.
\end{eqnarray}
This result is exact (to 1-loop in the quantum field theory)
for any $Z$ and and
$\epsilon \leq 1$, pair-production being energetically impossible
if $\epsilon > 1$.

One can now show that if $\delta \ll \sqrt{1 - \epsilon^2}$,
a good approximation to this integral expression for any $\epsilon
\leq 1$ is
\begin{equation}
\overline{\cal N}_{\rm spinor} \approx
\frac{(qE)^2 L}{4 \pi^3} \sqrt{\pi \delta} (1 - \epsilon^2)^{5/4}
e^{ -Z (1 - \sqrt{1 - \epsilon^2})}. \label{ii}
\end{equation}
For the particular subcase considered above, $\epsilon \ll 1$ and $\delta
\ll 1$, so long as $\delta \ll \epsilon^4$ one gets agreement with the
instanton-method approximation given  in Eq. (\ref{in res}) (after
multiplying that expression by the 2 spin states of the spinor).

In particular, for $\delta^2 \ll \epsilon^4 \ll \delta$, (where the first
subdominant term in the expansion of the exponent in (\ref{ii}) is large, but
not the second subdominant term), the approximation (\ref{ii}) to the exact
answer (\ref{i}) agrees with the 0-loop instanton-method approximation (\ref{in
res}) but not with the uniform-field approximation there, which is larger by
the factor $e^{\epsilon^2/(4 \delta)}$. However, for $\delta \ll \epsilon^4$,
then even the second subdominant term in the expansion of the exponent in
(\ref{ii}) is large in comparison with unity, so then the exact result is
significantly different from both the uniform-field approximation and the
0-loop instanton-method approximation (\ref{in res}), at least if (\ref{ii}) is
a good approximation to the exact result to all orders in $\epsilon$ when
$\delta \ll \sqrt{1 - \epsilon^2}$. However, we do obtain precisely (\ref{ii})
by properly calculating the 0-loop instanton action and making a gaussian
approximation for the integral over $\omega$ and ${\bf k}_{\perp}$.

In the case in which $\epsilon^2 \ll \delta \ll 1$, then both
approximations agree closely with the exact result. When we retain
the assumption that $\epsilon \ll 1$ but go to the opposite limit of
high maximum electric field strength, so $\delta \gg 1$, then the
exact integral expression (\ref{i}) reduces to the uniform-field
approximation of Eq. (\ref{un res}) (after that is multiplied by the
factor of 2), rather than to the 0-loop instanton-method
approximation also given there. That is, the 0-loop instanton-method
approximation gives a result that is too large by a factor of $3
\pi/8 \approx 1.178$. This is what we might expect, since the
potential is changing rapidly so the 0-loop instanton approximation,
using just the lowest-order (classical) action as we have done here,
would not necessarily be expected to be good in this limit. And yet
it is encouraging that it is only off by less than $20\%$ for the
pair-production rate.

While this paper was being revised, we have become aware of some
related papers that have recently appeared in the literature
\cite{Gies-Langfeld, LMG, MLG, Bast-Schubert, Gies-Klingmuller,
Dunne-Schubert, DGSW} which use the worldline method.  For a static
electric field in the $z$-direction that depends only on $z$, the
single-worldline instanton action $S_0$ is the same as the minimum
value of our 0-loop or classical instanton action $2S^{(0)}_{{\bf
k}_{\perp}}(\omega)$ as a function of the transverse momentum ${\bf
k}_{\perp}$ and the frequency $\omega$.  (The minimum is always at
${\bf k}_{\perp} = 0$, and in the special cases in which scalar
potential is an antisymmetric function of $z$, $A_0 (z) = -A_0
(-z)$, as in, e.g., the Sauter potential, the minimum is also at
$\omega = 0$.)

Dunne, Gies, Schubert, and Wang have now worked out the subleading
prefactor contribution \cite{DGSW} using the worldline instanton
method, which appears to agree with the gaussian approximation for
the integrals over $\omega$ and ${\bf k}_{\perp}$ in our approach
using $2S^{(0)}_{{\bf k}_{\perp}}(\omega)$.  For the Sauter
potential, both instanton methods give exactly the same expression
as Eq. (\ref{ii}) which we obtained as an approximation to the exact
double-integral Eq. (\ref{i}) from Nikishov \cite{nikishov}.

\section{Conclusion}

In conclusion, using the instanton method in the Coulomb
(space-dependent) gauge for constant electric and magnetic fields,
we have obtained new forms (\ref{b-p}) and (\ref{f-p})
for the pair-production rate in
scalar QED and spinor QED. These formulae agree
exactly with the well-known results.
Finally, we suggested a generalization of these formulae
to a constant magnetic field and an inhomogeneous electric field,
and compared the results to another method and to the exact answer for the
Sauter electric potential.

\acknowledgments

The authors thank G. Dunne, H. Gies, A. I. Nikishov, S. J. Rey, and C. Schubert
for useful discussions, and S.P.K. expresses his appreciation for the warm
hospitality of the Theoretical Physics Institute, University of Alberta. The
work of S.P.K. was supported by the Korea Science and Engineering Foundation
under Grant Nos. 1999-2-112-003-5 and R01-2005-000-10404-0, and the work of
D.N.P. by the Natural Sciences and Engineering Research Council of Canada.

\appendix
\section{Tunneling and Reflection Probability}

In the context of nonrelativistic quantum theory, we derive a
formula for the tunneling and reflection coefficients for bosons
under a general potential barrier.

Let the wave equation take the form
\begin{equation}
y'' (z) + Q(z) y(z) = 0, \quad Q = E - V(z),
\end{equation}
and have three different regions:
\begin{eqnarray}
\begin{cases} I = (-\infty, z_1) , &  Q > 0, \cr II = (z_1, z_2), &
Q < 0, \cr III = (z_2, \infty), & Q > 0. \cr \end{cases}
\end{eqnarray}
Here $z_1$ and $z_2$ are two turning points. We assume that an
incoming wave function from $- \infty$ in region (I) is partially
reflected back to $- \infty$ and partially tunnels through the
barrier in region (II) toward $\infty$ in region (III). Then the
wave function in region (I) is given by
\begin{equation}
y = y_I + B y_I^*, \label{qm I}
\end{equation}
where $y_I$ has the incoming flux toward the barrier, and in region (III)
\begin{equation}
y = C y_{III}
\end{equation}
(assuming an outgoing flux there, which actually corresponds to
an incoming group velocity for the relativistic waves of our paper).
Here the incoming flux is normalized to unity. The flux
conservation for equal momenta at $z = \pm \infty$ leads to the
relation
\begin{equation}
|B|^2 + |C|^2  = 1. \label{flux con}
\end{equation}
Then we may find a wave function of the form \cite{bender}
\begin{equation}
y_I = \exp \Bigl[\sum_{k = 0}^{\infty} S^{(2k)} \Bigr], \label{ben
wav}
\end{equation}
with ${\rm Im} (dS_0/dz) > 0$ to conform with the direction of
the incoming flux. The leading term in region (I) is given by
\begin{equation}
S^{(0)} (z) = i \int_{- \infty}^{z} \sqrt{Q(z)} dz,
\end{equation}
and in region (II)
\begin{equation}
S^{(0)} = i \int_{z_1}^{z_2} \sqrt{Q(z)} dz
\end{equation}
is what may be called the 0-loop or classical
action for the nonrelativistic problem.

We may analytically continue $y_I$ from region (I) to region (III)
through region (II) along a semicircle starting from $z_1$ and
ending at $z_2$ in the upper $z$-plane. Then $z \rightarrow e^{- i
\pi} z = -z$, and the leading term, for instance, takes the form
in region (III)
\begin{equation}
S^{(0)} = - i \int_{\infty}^z \sqrt{Q(z)} + i \int_{z_1}^{z_2}
\sqrt{Q(z)} + i \Biggl(\int_{- \infty}^{z_1} \sqrt{Q(z)} +
\int_{\infty}^{z_2} \sqrt{Q(z)}\Biggr). \label{lead III}
\end{equation}
In Eq. (\ref{lead III}), the first term in the right hand side has
the incoming flux, hence is the leading term of
$y_{III}^*$, and the second term is the instanton action, a real
quantity, through the barrier, and the last term gives a phase
factor. We may thus write the analytically continued wave function
as
\begin{equation}
y_I \rightarrow e^{S} (y_{III}^* + \alpha y_{III}).
\end{equation}
So the wave function in region (I) is analytically continued in
region (III) to
\begin{equation}
y_I + B y_I^* \rightarrow e^{S} (\alpha + B) y_{III} + e^{S} (1 +
B \alpha^*) y_{III}^*.
\end{equation}
As there is no incoming wave function in region (III), we have
\begin{equation}
e^{S} (\alpha + B) = C, \quad 1 + \alpha^* B = 0. \label{aux con}
\end{equation}
Solving Eq. (\ref{aux con}) together with Eq. (\ref{flux con}), we
finally obtain the coefficients for tunneling and reflection
\begin{equation}
|B|^2 = \frac{1}{ 1 + e^{-2S}}, \quad |C|^2 = \frac{1}{1 +
e^{2S}},
\end{equation}
which confirms the result in Refs. \cite{kim-page} and
\cite{froman}.


\begin{references}
\bibitem{footnote} There are different terms for $w = 2 {\rm Im} {\cal L}$,
which is not exactly the same as the pair-production rate. But since it agrees
to lowest order when small, Schwinger \cite{schwinger} used the term
`pair-production probability' for $w$, and Itzykson and Zuber \cite{itzykson}
used `pair-production rate'. Nikishov uses `pair-production' only for the
quantities of Eqs. (\ref{fer pair}) and (\ref{bos pair}). For a single mode,
the relation is given by Eq. (\ref{33}). In fact, $w$ describes the
vacuum-to-vacuum persistence or the decay rate of the vacuum.
\end{references}
\end{document}